\documentclass{article}

\usepackage{graphicx}
\usepackage{natbib}
\usepackage{amsmath}
\usepackage{amsfonts}
\usepackage[hmargin=2cm,vmargin={3.5cm,4cm}]{geometry}

\newtheorem{definition}{Definition}[section]

\title{aa}

\author{F. Polito, A. Petri, G. Pontuale, \& F. Dalton}
\title{Analysis of Metal Cutting Acoustic Emissions by Time Series Models}

\begin{document}

\maketitle

\begin{abstract}

	We analyse some acoustic emission time series
	obtained from a lathe machining process. Considering the dynamic evolution
	of the process we apply two classes of well known stationary stochastic time series
	models. We apply a preliminary root mean square (RMS) transformation
	followed by an ARMA analysis; results thereof are mainly related
	to the description of the continuous part (plastic deformation) of the signal.
	An analysis of acoustic emission, as some previous works show,
	may also be performed with the scope of understanding the evolution of the ageing
	process that causes the degradation of the working tools.
	Once the importance of the discrete part of the acoustic
	emission signals (i.e. isolated amplitude bursts) in the ageing process
	is understood, we apply a stochastic analysis based on point processes
	waiting times between bursts and to identify a parameter
	with which to characterise the wear level of the working tool. A Weibull
	distribution seems to adequately describe the waiting times distribution.

\end{abstract}


\section{Introduction}

	When a solid body is subjected
	to a varying stress, acoustic waves (i.e. pressure
	waves propagating inside
	of the body),
	often reaching very high frequencies, are generated.
	Examples of this phenomenon are the noise
	produced by hitting a metal block
	with a hammer
	or the creaking of a wooden floor; the waves propagate
	inside and on the surface of the materials
	before dissipating in the surrounding gaseous medium.
	The process of generating \emph{acoustic waves}
	in stressed materials is called
	\emph{acoustic emission} (AE).
	The acoustic emission can be
	recorded by means of a
	transducer (i.e. a sensor
	in contact with the solid body which transforms
	the elastic waves into electric signals).
	Usually the emission of acoustic waves
	may also be associated with microfractures
	inside the solid or, in general,
	a degradation of its condition.
	Therefore an analysis of the
	acoustic emissions can reveal
	the level of degradation of the solid.

	In particular we note
	that the study of acoustic emission
	is developing in the field of
	\emph{tool condition monitoring} (TCM)
	where it is important, for example to avoid
	damaging machinery and
	to maximise productive capacity.
	AE analysis is an easily implemented
	and economical technique and it allows
	real-time monitoring of
	working tool conditions.

	Other statistical analyses have been
	previously conducted
	using the same experimental setup described
	in this article
	\citep{farrelly:metal,petri:tcm2,petri:tcm}.
	Results demonstrated that acoustic
	emission analysis is particularly relevant
	for the study of the ageing of work tools
	and the authors elucidated methods
	for the elicitation and reconstruction
	of the pdf of the root mean square amplitude signal.

	In this study we analyse some specific high frequency
	acoustic emissions by means of time series
	statistical-probabilistic models:
	ARMA models \citep{box:time,batt:prev} and
	models for point processes \citep{batt:prev,cox:stoch}.
	An objective of
	this study is to assess their suitability
	for explaining AE phenomena, and to estabilish if it might be
	possible to implement a tool wear monitoring algorithm.

\begin{figure}
		\centering
		\includegraphics[scale=1]{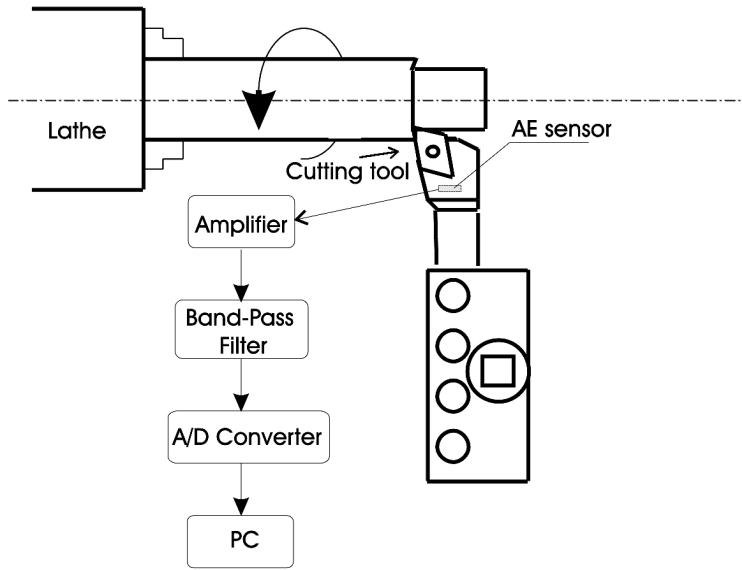}
		\caption{\label{figure_zero}The acoustic emissions were recorded by means of a
			transducer, preamplified, filtered, digitized and then stored.}
	\end{figure}

	The experimental setup (presented in figure~\ref{figure_zero})
	consisted of
	a mechanical lathe working on stainless
	steel bars and a transducer with which the resulting acoustic
	emissions were recorded.
	This signal was preamplified and filtered
	by a band-pass filter to isolate the frequencies of interest.
	The digitization was performed by means of a digital
	oscilloscope with a sample frequency of $f_0= 2.5$ Mhz.
	The cutting speed ranged from $0.5$ to $1$ ms$^{-1}$
	and the cutting depth was $2$
	mm.

	Three types of working tools were used
	for the analyses: new, partially and
	totally worn. For the new and totally
	worn tools we conducted one acquisition
	with each of eight different tools. For the
	partially worn tools only four acquisistions were
	conducted.
	For each acquisition we recorded $15$
	AE time series, each composed of
	$40960$ consecutive points (i.e. $614400$ points
	for each acquisition). The time duration of a
	single time series is $16.38$ ms.

	All analyses were conducted with
	the R environment for statistical computing (\cite{r}, http://www.R-project.org),
	a very powerful
	programming environment mainly used for statistical
	data analysis and modelling.

\section{Data collected and preliminary analyses}

\begin{figure}
		\centering
		\includegraphics[scale=0.45]{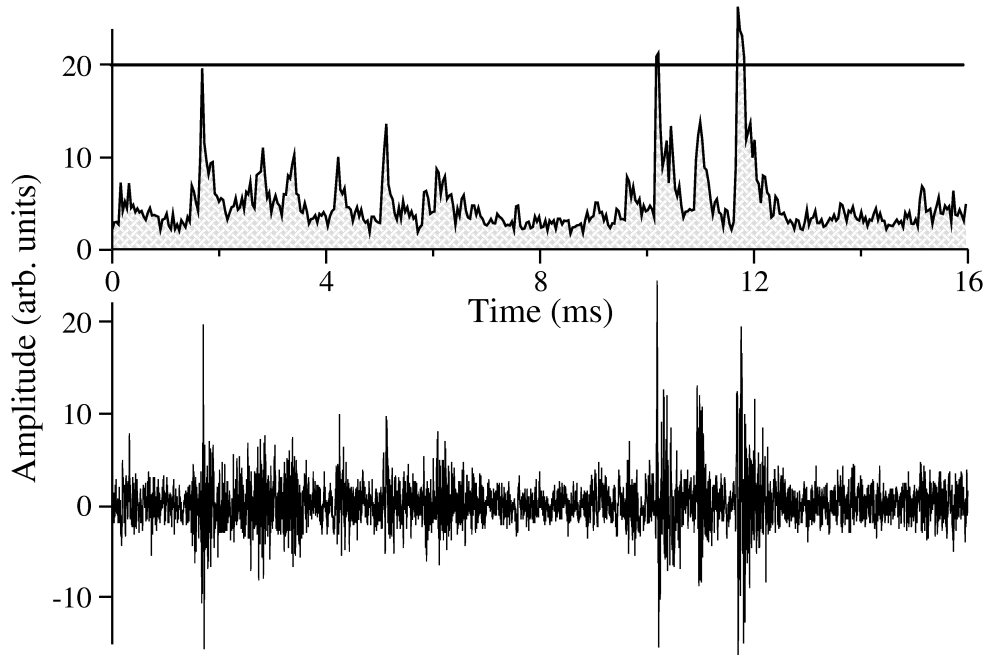}
		\caption{\label{figure_one}A typical AE signal with its RMS transform above
			with a threshold at level 20.}
	\end{figure}

	When we take a close look at the collected
	time series that result from the experiments,
	it is possible to distinguish two different
	and superimposed parts of the signal.
	As we can see in the lower part of figure~\ref{figure_one},
	there is a \emph{continuous} part characterized by
	a relatively constant variability (essentially due
	to plastic deformation of the material)
	and a more interesting part composed of bursts of
	different and high amplitudes (usually associated
	with microfractures in the tool or with material splinters striking
	it).

	The natural time scale of the cutting process and the
	short duration of each time series
	suggest that each time series will be effectively stationary;
	this is verified by the Dickey-Fuller
	test \citep[par. 18.3.3]{greene:ecan}.

	Because of the large amount of data (due to the
	high sample rate necessary to capture the highest
	frequencies), it is necessary to operate a
	\emph{data reduction} by means of a transform that
	has some physical meaning \citep{kannatey:wear}.
	\begin{definition}[RMS Values]
		Let $x_s$, $s \in \mathbb{N}$ be a time series taking values in $\mathbb{R}$ and
		$T$ the number of samples in some interval over which the RMS $y_t$
		is to be calculated. We define
		\begin{equation}
			y_t = \sqrt{ \frac{ \sum_{j=1}^{T} \
			x_{ \left( t-1 \right) T+j}^2 }{ T } }, \qquad t \in \mathbb{N}.
		\end{equation}
	\end{definition}

	If a time series is of length $N$ (multiple of $T$), then the RMS series $y_t\in \mathbb{R}$
	has $N/T$ points, each proportional to the acoustic energy emitted in the interval $(tT-T,tT]$.
	Our choice of $T$ is motivated by examination of the AE spectral density function $\tilde{x}(f/f_0)$,
	which is naturally separable into lower and higher frequency regions by an almost zero density
	region centred at $f/f_0\sim 5\times 10^{-3}$.
	Furthermore, analysis of $\tilde{x}(f/f_0)$ on a moving window
	demonstrates that the low frequency region is largely constant while the higher part depends
	critically on the presence or absence of bursts in the window.

	As bursts are typically separated by the O(10$^4$) samples, we conclude
	that the lower part of the spectrum is due to the process generating the series of bursts,
	whereas the higher part is due both to the dynamics of individual bursts and to the structure
	of the continuous part of the AE signal. In order to obtain a meaningful RMS, that is, to reduce
	the data to a small number of points characterising the AE noise level, we chose $T=100$
	corresponding to the spectral gap at
	$f/f_0=5\times 10^{-3}$ (equal to a physical interval of
	$40\mu s$).

\section{Results for ARMA models}

	We now subject this RMS series $y_t$ to analysis to determine if some feature of the
	resulting model can be exploited to estimate the wear level.
	ARMA (Auto Regressive Moving Average) linear models are widely used for modelling
	stationary time series in general
	\citep*{brock:theory,pries:spectral,shum}
	and are very flexible with respect to real-world applications.
	\citet{76377} have successfully discriminated normal and pathological
	patients using ARMA analyses on acoustic signals from the respitory tract.
	\citet{Hol} have used a more generalised ARMA model to forecast failure in
	mechanical systems and \citet{freq} have presented a review of acoustic applications
	of ARMA models, defining in detail a protocol for the analysis of acoustic spectral
	features.  Some other examples may be taken from the fields of medicine \citep*{cor}, finance
	\citep*{Gha,wen}, languages \citep*{paw} and engineering (\cite{kannatey:wear},
	\cite{freq2}, \cite{wind}, cite{ground},
	\cite{ground2}, \cite{Na}).
	Previous research on machining by means of techniques related to time series analysis
	was conducted by Professor Wu and his team, and reported in several papers dating back to the
	late 1980s.
	Amongst other we mention \citet{Kim1989282}, \citet{Fassois1989153}, \citet{Yang1985336},
	\citet{Ahn198591} and \citet{1989MSSP}.

	Although few naturally occurring processes
	are intrinsically linear, the aim of
	this section of the analysis is to understand if
	ARMA models are suitable for the purpose
	of representing acoustic emission signals,
	and if a linear approximation could give us
	enough information about the dynamical process itself.
	\begin{definition}[ARMA processes]
		Let us consider a real valued process $X_t, t \in \mathbb{N}$. It is called an
		$\text{ARMA} \left(p,q \right)$ process (combining an $AR \left( p \right)$ and
		$MA \left( q \right)$ model) if
		\begin{equation}
			\label{arma}
			X_t=c_0+ \phi_1 X_{t-1} + \cdots + \phi_p X_{t-p} + \epsilon_t + \theta_1 \epsilon_{t-1}
			+ \cdots + \theta_q \epsilon_{t-q}
		\end{equation}
		where $c_0$ is the intercept, $p$ and $q$ are
		respectively the number of parameters in the autoregressive and moving
		average part of the process. The $p + q$ parameters $\phi_1 \cdots \phi_p,\theta_1 \cdots
		\theta_q$ must be chosen such that $|\phi_i|<1,i \in \{ 1,\cdots, p\}$ and $|\theta_j|<1,
		j \in \{1,\cdots,q \}$ to ensure stationarity and invertibility of the process. The
		process $\epsilon_t$ called innovations is taken to be a white noise (see e.g. the
		classical text of \cite{box:time} or \cite{brock:theory}
	\end{definition}

\begin{figure}
		\makebox{\includegraphics[scale=0.27]{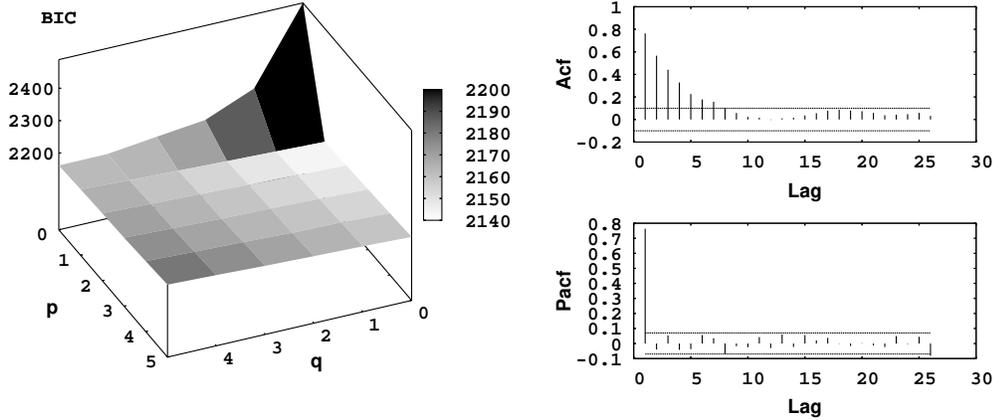}}
		\caption{\label{figure_two}BIC matrix and autocorrelation functions.}
	\end{figure}

	For each level of wear and for each recorded
	time series we have conducted a full analysis following
	the Box-Jenkins iterative procedure \citep*{box:time}.
	The resulting best model is
	generally very simple;
	often it has only three or four parameters.
	We can analyse the results, for example, for
	an RMS time series taken with a fully
	worn tool.
	At the right side of figure~\ref{figure_two}
	we can see the autocorrelation function
	and the partial autocorrelation function
	of the series. Their shapes (the acf's
	decay is exponential and the pacf is
	zero after lag one) and an analysis
	of the spectral density function suggest a simple
	AR$\left(1\right)$.
	Furthermore if we apply the procedure in a
	completely automated manner \citep[par. 9.3]{brock:theory},
	the resulting best model is indeed
	an AR$\left(1\right)$ model.
	\begin{align}
		\label{model}
		X_t & = \hat{c_0} + \hat{\phi} X_{t-1} + \hat{\epsilon}_{t} \nonumber \\
		    & = 7.5499 + 0.7632 X_{t-1} + \hat{\epsilon}_{t}
	\end{align}
	where $\hat{c_0}$ is the estimated intercept, $\hat{\phi}$ is the sole estimated autoregressive
	parameter and $\hat{\epsilon}_t$ are the residuals (i.e. the estimated innovations).
	All the parameters are statistically significant.
	\begin{definition}[BIC -- Bayesian Information Criterion]
		We define the Bayesian Information Criterion for a model with $(p+q)$ parameters
		and $N$ observations as
		\begin{align}
			BIC \left( p,q \right) & = N \log \hat{\sigma}^2 \left( p,q \right) +
			\left( p+q \right) \log N + \nonumber \\
			& - \left( N-p-q \right) \log \left( 1-\frac{p+q}{N} \right)
			+ \left( p+q \right) \log \left( \frac{1}{p+q} \left( \frac{\hat{\sigma}^2_*}{
			\hat{\sigma}^2 \left(p,q \right)} \right) \right)
		\end{align}
		where $p$ is the number of parameters in the autoregressive part of the
		model, $q$ is the number of parameters in the moving average part, $\hat{\sigma}^2
		\left(p,q \right)$ is the residuals variance calculated after having fitted an
		$ARMA \left(p,q \right)$ model and $\hat{\sigma}^2_*$ is the sample variance of
		the observations (for details see \cite{pries:spectral} page 375).
	\end{definition}
	To better understand the behaviour of the BIC it is useful to give the following representation
	also presented in \cite{pries:spectral}
	\begin{equation}
		BIC \left( p,q \right) = N \log \hat{\sigma}^2 \left( p,q \right) +
			\left( p+q \right) \log N + o \left( p+q \right)
	\end{equation}
	The BIC must thus be minimised w.r.t. $(p,q)$ in order to select the model which best
	explains the observations with the minimum of parameters.

	At the left side of figure~\ref{figure_two} we can
	see the levelplot for the BIC matrix.
	For each combination of order $\left(p,q\right)$
	of the AR and MA parts, maximum likelihood
	estimates for the parameters are computed and
	BIC$\left(p,q\right)$ is recorded on the matrix.
	The minimum BIC value indicates that the best
	model is that of eq.~(\ref{model}).

	Following the usual procedure to validate the
	model, we conduct an analysis of \emph{whiteness}
	on the residuals. The autocorrelation and
	partial autocorrelation functions for the
	estimated residuals are both within the white noise
	confidence band for
	lags strictly positive, and
	the spectral density function is uniform.
	Whiteness tests in both the time and frequency
	domains (Ljung-Box for various lags and
	cumulative periodogram tests) were conducted with
	positive outcome. This model therefore explains the
	whole linear dependence between the
	variables in the process.

	At this point we have a model which adequately
	describes the stationary part of the AE RMS
	signal.
	Furthermore we can see that the mean
	of the residuals' variance
	decreases with
	increasing tool wear.
	However the decrease of the residuals'
	variance (with respect to the wear level)
	is not due to the better explanation
	of the data by the model, but instead
	to the decreasing number of bursts in the time series.
	Therefore ARMA models are more suitable for the description
	of the essentially continuous transformed signal part
	(the one due to plastic deformation).

	\begin{figure}
		\centering
		\makebox{\includegraphics[scale=0.5]{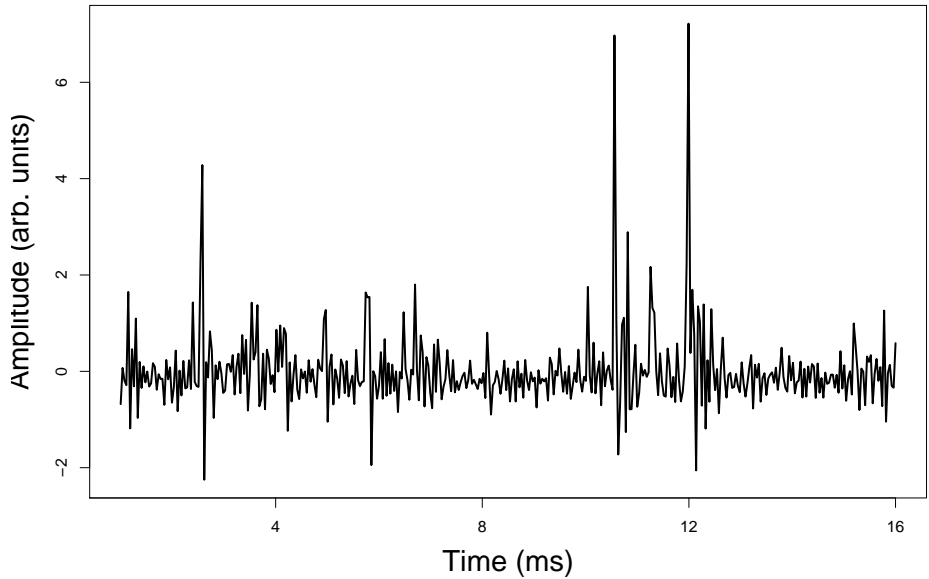}}
		\caption{\label{res}A typical residuals time series after adapting an ARMA model. We note that bursts are still present.}
	\end{figure}

\section{Point processes}\label{popro}

	Taking a closer look at the residuals (figure \ref{res}), we
	note that bursts are still present, though
	with a smaller amplitude.
	Therefore the bursts are not explained by the AR(1) model
	of the preceding section, and therefore not by any $\text{ARMA}(p,q)$.
	This indicates a decrease in the mean burst frequency
	with increasing wear level.
	If we wish to understand the evolution of the
	acoustic emissions with respect to the
	wear level of the tool, we must take
	into account the dynamical and statistical
	process that generates the bursts.

\begin{figure}
		\centering
		\makebox{\includegraphics[scale=0.45]{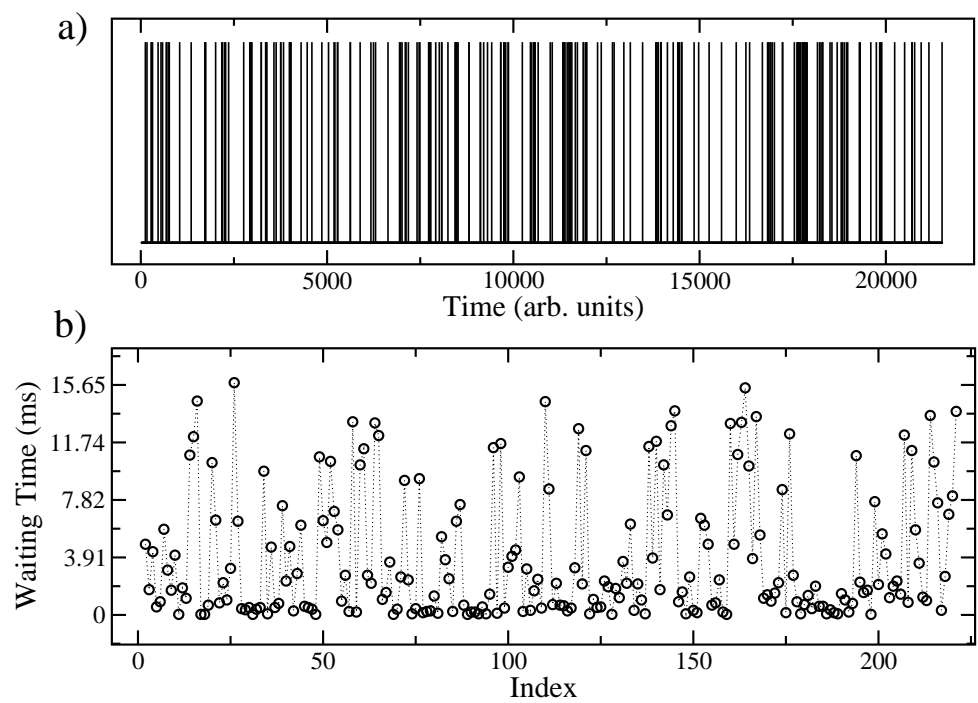}}
		\caption{\label{figure_four}a) The sequence of burst events as a function of time. b) The waiting time between events indexed by their order in the sequence of bursts.}
	\end{figure}

	The acoustic emission bursts are usually
	associated with micro-fractures in the
	work tool. They are, effectively, singular
	events (the
	exponential decay is due to
	the transducer response). Furthermore the number of bursts
	seems to change with the wear level.
	Point processes are a tool widely used in
	modelling inherently point phenomena
	(see e.g. \cite{Paparo}, \cite{telo}).
	In this section we consider the bursts
	as the outcome of
	a point process and try to understand
	the behaviour of this process as
	the wear level increases.
	Figure \ref{figure_four} shows
	the overall point process for
	new tools and the observed waiting
	times between bursts. The waiting times
	between events are registered and presented
	in figure~\ref{figure_four}(b) indexed by their order in the sequence
	of bursts.

	The identification of the events was
	performed by placing a threshold
	at various levels to obtain information
	about bursts of different amplitudes.
	In particular we chose four levels of
	amplitude $40$, $50$, $60$ and $70$.
	These thresholds are expressed in arbitrary units
	proportional to the signal's amplitude
	which depends on the data collection chain.
	For an example see fig.~\ref{figure_one}
	where the placement of
	a threshold at level 20 identifies
	two burst events in the RMS-transformed
	time series.
	The waiting times process is then
	calculated for all the thresholds
	considered and is in effect a renewal
	process \citep{cox:renewal}. It is
	possible to verify
	the lack of correlation
	(second order dependence) in the
	waiting times process by calculating
	the spectral density function and
	the complete and partial
	autocorrelation functions.

	The hypothesis of
	exponentially distributed waiting times
	(Poisson process) and Pareto (power-law)
	distributed waiting times (fractal point
	process) as in \citet{thurner:fracpoin}
	and \citet{lowen:fract}
	is addressed conducting Kolmogorov-Smirnov
	goodness of fit tests on the p.d.f. for
	all thresholds and wear levels.
	The tests resulted not significant,
	so another hypothesis for the distribution must be
	proposed.

	As stated in \citet{malevergne:weitopar},
	the Weibull distribution
	(also known as \emph{Stretched Exponential})
	can have (for typical parameters' values)
	both the features of an exponential distribution
	and a Pareto distribution:
	\begin{equation}
		f \left( x \right) = \frac{ \alpha}{ \beta} \left( \frac{x}{ \beta} \right)^{ \alpha -1} e^{ - \left( \frac{x}{ \beta
			} \right)^{ \alpha}}
	\end{equation}
	where $\beta$ and $\alpha$ are, respectively, the scale and the
	shape parameter ($x \geq 0, \alpha,\beta > 0$).

	When $\alpha = 1$ a Weibull is
	an exponential distribution with rate
	$\lambda = \frac{1}{ \beta}$. This kind of
	distribution, therefore, allows us to describe
	point processes that have waiting times following
	an \emph{exponential behaviour}, but that can exhibit
	a heavier tail.
	\begin{table}
		\caption{\label{results}Maximum Likelihood Estimates and P-Values of Kolmogorov-Smirnov Tests}
		\centering
		\fbox{%
		\begin{tabular}{*{9}{c}}
			\multicolumn{1}{c}{Thresh.} & \multicolumn{3}{c}{New} & \multicolumn{3}{c}{Worn} & \multicolumn{2}{c}{P-Value} \\
			& $\hat{\alpha}$ & $\hat{\beta}$ & $ \hat{ \mu} $ & $\hat{\alpha}$ & $\hat{\beta}$ & $ \hat{ \mu} $ & New & Worn \\
			\hline
			70 & 0.845 & 98.53 & 90.92 & 0.764 & 138.98 & 124.56 & 0.9907 & 0.3494 \\
			60 & 0.875 & 119.68 & 111.95 & 0.786 & 138.99 & 125.38 & 0.5507 & 0.2388 \\
			50 & 0.822 & 109.99 & 100.53 & 0.942 & 148.79 & 144.06 & 0.606 & 0.1154 \\
			40 & 0.754 & 83.1 & 74.27 & 0.907 & 138.90 & 132.02 & 0.3204 & 0.1055 \\
		\end{tabular}}
	\end{table}
	Table~\ref{results} summarises the
	maximum likelihood estimates for the Weibull
	parameters $\hat{\alpha}$ and $\hat{\beta}$.
	We have calculated the estimates for each threshold
	and each wear level, finding the estimated mean
	$\hat{\mu}$ and performed a Kolmogorov-Smirnov
	test to see if the Weibull hypothesis can hold.
	As can be seen from the P-values in the table, the test is
	not significant for each level and each threshold.
	Table~\ref{results} contains only the
	estimates for the \emph{new} and \emph{totally
	worn} tools.
	The half worn tool has sufficient data for this
	analysis only with the lowest threshold;
	the result is
	consistent with the other wear levels (i.e. $\hat{\mu}=113.15$).

\begin{figure}
		\centering
		\includegraphics[scale=0.45]{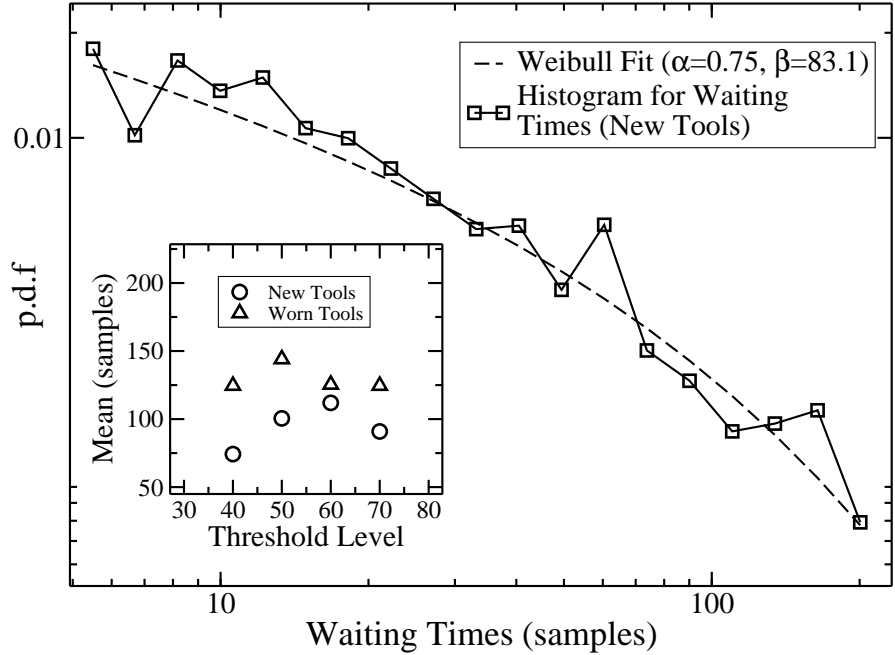}
		\caption{\label{figure_three}Weibull fit for the pdf of the waiting times process. The mean
		inter-burst waiting time consistently increases with increasing tool wear.}
	\end{figure}

	Figure~\ref{figure_three} shows
	the maximum likelihood Weibull fit for the distribution
	of waiting times between the bursts (new tools,
	threshold equal to $40$). The estimated parameters
	are $\hat{\alpha} = 0.75$ and $\hat{\beta} = 83.1$.
	In the inset we note that, for all thresholds considered,
	the estimated
	mean of the waiting times increases with the tool wear level.
	For purposes of TCM we could therefore monitor the mean
	of the estimated distribution to decide whether
	is necessary to change the tool or not.

\section{Conclusions}

	In this analysis we have shown an application of time series
	models to acoustic emission signals from metal cutting
	processes.
	Before this work few stochastic analyses
	were performed on this particular type of data.
	After transforming them by means of the RMS transform,
	we applied standard time series statistical
	techniques to explain the underlying process
	that generates the phenomenon under consideration.
	In particular initially we used linear ARMA models together
	with the well known Box-Jenkins iterative procedure.
	We found that these linear models with
	a small number of parameters are suitable for
	the description of the linear contribution
	of the background part of the signal. The variance
	of the residuals decreases when the wear level increases
	but this effect is due mainly to the decreasing
	number of bursts in the time series. Even though,
	for the purposes of TCM, we could in principle
	monitor that variance, we have obtained better results
	looking directly at the underlying mechanism
	generating the burst process.

	A renewal point process with Weibull distributed
	waiting times seems to represent the burst
	process very well. In particular, for each
	threshold and each wear level the
	Kolmogorov-Smirnov test results not significant.
	When the tool wear level increases the
	tail of the distribution becomes heavier
	and the estimated mean of the
	waiting times process increases.
	The number of bursts is actually
	correlated with increasing
	tool wear, therefore we could
	identify the estimated mean as the
	parameter that should be used to
	monitor the condition of the
	working tool (the tool will be
	substituted when the estimated
	mean becomes sufficiently high).

	In conclusion, it is important to
	underline that, from a methodological
	point of view, both ARMA models
	and Weibull point processes are suitable for
	modelling the phenomenon (even though
	ARMA models are somewhat limited to
	explaining the linear contribution of the
	plastic deformation).
	What appears of importance is that
	the shape of the distribution of
	waiting times changes
	in the sense that the tail becomes
	heavier when the wear level increases.
	Therefore the process creating the bursts
	seems to evolve in a fundamental manner.

	In future work, it would certainly be very interesting to
	follow the evolution of the distribution shape
	along the whole life of the cutting tool to
	try to understand better the dynamical
	properties of the underlying process.

	\vspace{.5cm}
	\textit{Acknowledgement:} The authors wish to thank the referees for their
	helpful comments and insights.
\nocite{*}
\bibliographystyle{Chicago}
\bibliography{art}

\end{document}